\documentstyle[12pt]{article}
\begin{document}

\title{\Large\bf Axial and gauge anomalies in a theory with one and
two-form gauge fields}

\author{\\
R. Amorim\thanks{\noindent e-mail: amorim@if.ufrj.br}~ and J.
Barcelos-Neto\thanks{\noindent e-mail: barcelos@if.ufrj.br}\\ 
Instituto de F\'{\i}sica\\
Universidade Federal do Rio de Janeiro\\ 
RJ 21945-970 - Caixa Postal 68528 - Brasil\\}
\date{}

\maketitle
\abstract
We study the problem of axial and gauge anomalies in a reducible
theory involving vector and tensor gauge fields coupled in a
topological way. We consider that vector and axial fermionic currents
couple with the tensor field in the same topological manner as the
vector gauge one. This kind of coupling leads to an anomalous axial
current, contrarily to the results found in literature involving
other tensor couplings, where no anomaly is obtained.

\vfill\noindent
\vspace{1cm}

\newpage
\section{Introduction}

\bigskip
The interest for tensor gauge fields dates back to more than twenty
years. They were considered by Kalb and Ramond \cite{Kalb} with the
motivation that they could carry the force among string interactions.
At the same time, Cremmer and Scherk \cite{Cremmer} considered these
fields coupled in a topological way with the usual vector gauge one,
with the purpose in obtaining a kind of dynamical breaking of the
gauge symmetry and a consequent mass generation for the vector field.
This same problem has been considered nowadays in a version where the
mass generation is carried out as an effective theory when the tensor
field is conveniently eliminated \cite{Allen,Barc1,Barc2}. We mention
that antisymmetric tensor fields also appear as one of the massless
solutions of string theories, in company with photons, gravitons etc.
\cite{Green}

\medskip
It is also opportune to mention that the particular structure of
constraints involving tensor fields is an interesting subject for
its own rights. They constitute a natural example of reducible
theory, in a sense that the first-class constraints \cite{Dirac} are
not all independent. Many developments have been done in this
direction too \cite{Kaul,Henneaux}.

\medskip
Our purpose in this paper is to study the problem of anomalies, where
the fermionic vector and axial currents also couple to the tensor
field.  We consider that this coupling has the same topological
nature of the vector-tensor gauge ones \cite{Cremmer}. It is
important to emphasize that this differs from the usual tensor
coupling that appear in the literature, where the tensor field
(considered as an external field) couples with a tensor current
\cite{Clark}. Using the Fujikawa path integral formalism
\cite{Fujikawa}, we show that this topological coupling leads to a
contribution for the axial current anomaly. This result is new
comparing with the ones found in literature where no contribution for
the axial current anomaly is found. These developments are done at
Sec. 2.

\medskip
In Sec. 3, we  show that the $U(1)$ and the tensor gauge symmetries
are not obstructed in the considered model. It is interesting to
note, however, that if one considers chiral couplings between the
vector and tensor fields with the fermionic currents, the $U(1)$
gauge symmetry is obstructed due to anomalies. Here, to perform these
calculations, we use the field-antifield formalism \cite{BV1}, the
best known method to treat reducible theories in a covariant way.  As
in the usual axial current anomaly case, the contribution of the
tensor sector to the $U(1)$ gauge anomaly is not trivial. However it
keeps the form of a total derivative times the $U(1)$ ghost, when a
convenient regularization is adopted.

\vspace{1cm}
\section{Axial current anomaly}
\renewcommand{\theequation}{2.\arabic{equation}}
\setcounter{equation}{0}

\bigskip
Let us start from the action involving vector and tensor gauge
fields \cite{Cremmer}

\begin{equation}
S=\int \,d^4x\,\left[-\,\frac{1}{4}\,F_{\mu\nu}F^{\mu\nu}
+\frac{1}{12}\,H_{\mu\nu\rho}H^{\mu\nu\rho}
+\frac{1}{2}\,m\,\epsilon_{\mu\nu\rho\lambda}\,
A^\mu\partial^\nu B^{\rho\lambda}\right]
\label{2.1}
\end{equation}

\bigskip\noindent
where $F_{\mu\nu}$ and $H_{\mu\nu\rho}$ are totally antisymmetric
tensors written in terms of the potentials $A_\mu$ and $B_{\mu\nu}$
(also antisymmetric) through the curvature tensors

\begin{eqnarray}
F_{\mu\nu}&=&\partial_\mu A_\nu-\partial_\nu A_\mu
\nonumber\\
H_{\mu\nu\rho}&=&\partial_\mu B_{\nu\rho}
+\partial_\nu B_{\rho\mu}+\partial_\rho B_{\mu\nu}
\label{2.2}
\end{eqnarray}

\bigskip\noindent
We notice that the vector and tensor gauge fields are coupled in a
topological way. It is a well known fact that the system represented
by $S$ is invariant under the gauge transformations

\begin{eqnarray}
\delta A_\mu&=&\partial_\mu\Lambda
\label{2.3a}\\
\delta B_{\mu\nu}&=&\partial_\mu\Lambda_\nu-\partial_\nu\Lambda_\mu\,.
\label{2.3b}
\end{eqnarray}

\bigskip\noindent
Although (\ref{2.3a}) is the usual irreducible $U(1)$ gauge symmetry,
(\ref{2.3b}) is reducible, since $\delta B_{\mu\nu}$ vanishes
identically if the vector parameter is the gradient of a scalar. At
quantum scenario the symmetries (\ref{2.3a}) and (\ref{2.3b}) are not
obstructed.  Integrating out the tensor fields leads to a non-local
$U(1)$ gauge invariant but massive effective vector theory
\cite{Barc1}.

\medskip
Let us now introduce matter field in this theory. We consider that
the fermionic vector current also has a topological coupling with the
tensor field, with action given by

\begin{equation}
S_0=\int d^4x\,\left[-\,\frac{1}{4}\,F_{\mu\nu}F^{\mu\nu}
+\frac{1}{12}\,H_{\mu\nu\rho}H^{\mu\nu\rho}
+\frac{1}{2}\,m\,\epsilon_{\mu\nu\rho\lambda}\,
A^\mu\partial^\nu B^{\rho\lambda}
+i\,\bar\psi\slash\!\!\!\!D\psi\right]
\label{2.3}
\end{equation}

\bigskip\noindent
where $D_\mu$ is a covariant derivative that also contains tensor
gauge fields, 

\begin{equation}
D_\mu=\partial_\mu+ieA_\mu+\frac{1}{2M}\,
\epsilon_{\mu\nu\rho\lambda}\,\partial^\nu B^{\rho\lambda}
\label{2.4}
\end{equation}

\bigskip\noindent
The parameter $1/M$ that appears in the Eq. (\ref{2.4}) is the
coupling between  $B_{\mu\nu}$ and the vector current. This kind of
coupling means that the theory described by Eq. (\ref{2.3}) is
nonrenormalizable. 

\medskip
In this section, we consider the axial current anomaly by using the
Fujikawa path integral technique \cite{Fujikawa}. As it was already
previously mentioned, we emphasize that the study of anomaly
involving tensor couplings that is found in literature differs from
the one we are going to develop here. Usually, one considers the
tensor field as an external field and coupled to a tensor current
\cite{Clark}. We notice that in our case, the tensor field has
dynamics and is coupled in a topological way to the same vector 
current coupled to the vector potential. 

\medskip
The axial current anomaly arises in the Fujikawa approach from the
fact that the measure $[d\bar\psi][d\psi]$ is not invariant under the
chiral gauge transformations

\begin{eqnarray}
\psi(x)&\longrightarrow&\psi^\prime(x)
=e^{i\epsilon(x)\gamma_5}\,\psi(x)
\nonumber\\
\bar\psi(x)&\longrightarrow&\bar\psi^\prime(x)
=\bar\psi(x)\,e^{i\epsilon(x)\gamma_5}
\label{2.4a}
\end{eqnarray}

\bigskip\noindent
It can be shown that \cite{Fujikawa} 

\begin{equation}
[d\bar\psi][d\psi]=[d\bar\psi^\prime][d\psi^\prime]\,
\exp\,{2ie\int d^4x\,\epsilon(x)\,I(x)}
\label{2.4b}
\end{equation}

\bigskip\noindent
for infinitesimal transformations $\epsilon(x)$. $I(x)$ is a
divergent quantity given by

\begin{equation}
I(x)=\sum_n\phi^\dagger_n(x)\,\gamma_5\,\phi_n(x)
\label{2.4c}
\end{equation}

\bigskip\noindent
where $\phi_n(x)$ is an orthonormal set of eigenfunctions of some
hermitian operator. This quantity needs to be regularized. We use the
operator (\ref{2.4}) (conveniently Wick rotated to an hermitian form)
in order to do so. It is not necessary to go into details to do this.
We can just consider the final result given in literature
\cite{Fujikawa} and make the replacement

\begin{equation}
A_\mu\longrightarrow\tilde A_\mu=A_\mu
-\,\frac{i}{2eM}\,\epsilon_{\mu\nu\rho\lambda}\,
\partial^\nu B^{\rho\lambda}
\label{2.5}
\end{equation}

\bigskip\noindent
Since the generating functional must be independent of the parameter
$\epsilon(x)$, we thus obtain

\begin{equation}
\partial_\mu j^\mu_5=\frac{e^2}{16\pi^2}\,
\epsilon^{\mu\nu\rho\lambda}\,
\tilde F_{\mu\nu}\tilde F_{\rho\lambda}
\label{2.6}
\end{equation}

\bigskip\noindent 
where $\tilde F_{\mu\nu}$ is the field strength defined in terms of
the $\tilde A_\mu$. The combination of (\ref{2.5}) and (\ref{2.6})
gives

\begin{eqnarray}
\partial_\mu j^\mu_5&=&\frac{e^2}{16\pi^2}\,
\epsilon^{\mu\nu\rho\lambda}\,F_{\mu\nu}F_{\rho\lambda}
+\frac{1}{24\pi^2M^2}\,\epsilon^{\mu\nu\rho\lambda}\,
\partial^\alpha H_{\mu\nu\lambda}\partial^\beta H_{\alpha\mu\beta}
\nonumber\\
&&\phantom{\frac{e^2}{16\pi^2}\,
\epsilon^{\mu\nu\rho\lambda}\,F_{\mu\nu}F_{\rho\lambda}}
-\frac{ie}{8\pi^2M}\,F_{\mu\nu}\partial_\rho H^{\rho\mu\nu}
\label{2.10}
\end{eqnarray}

\bigskip\noindent
We notice in the relation above the contribution for the axial
current anomaly originated from the tensor coupling we have
considered. These terms can also be written as a total derivative as
it occurs in the usual anomalous case. 

\vspace{1cm}
\section{Gauge anomalies}
\renewcommand{\theequation}{3.\arabic{equation}}
\setcounter{equation}{0}

\bigskip
In the preceding section we have analyzed the anomalous divergence of
the fermionic chiral current when both vector and tensor fields are
coupled to non-chiral fermions.  It is interesting to argue if
there are quantum obstructions to the gauge symmetries presented by
action (\ref{2.3}). As the gauge symmetries associated to the tensor
sector are reducible, and also because we are interested in
keeping covariance at each stage, it is useful to search for gauge
anomalies with the aid of the field-antifield formalism \cite{BV1}.
The case involving only tensor fields can be found in
References \cite{Henneaux,Gomis}. The case where vector and tensor
fields are topologically coupled was considered in \cite{Barc1}. The
inclusion of fermions induces only simple modifications regarding the
results found in \cite{Barc1}. We get for the field-antifield action

\begin{eqnarray}
\bar S&=&S_0+\int d^4x\,
\Bigl(i\,A^\ast_\mu\,\partial^\mu c+\bar c^*\,b
-ie\,\psi^\ast c\psi+ie\bar\psi c\bar\psi^\ast
+i\,B^\ast_{\mu\nu}\,\partial^\mu d^\nu
\nonumber\\
&&\phantom{S_0+\int d^4x\,\Bigl(i\,A^\ast_\mu\,\partial^\mu c}
+d^\ast_\mu\,\partial^\mu d+\bar d^\ast_\mu\,e^\mu
+i\,\bar d^\ast\,\bar f+i\,\eta^*f\Bigr)
\label{3.1}
\end{eqnarray}

\bigskip\noindent 
where $S_0$ is given by (\ref{2.3}). In the expression above we have
introduced the gauge fixing term for the vector gauge and Dirac
fields, consisting of ghosts, trivial pairs and corresponding
antifields, which essentially represent the sources for the BRST
transformations of the field sector. We have also considered the
gauge fixing for the tensor field that is a bit more involved due to
its reducibility. It was demanded the introduction of ghosts for
ghosts and the corresponding antifields, besides trivial pairs for
the implementation of the gauge fixing. For completeness, let us
introduce the parities and ghost numbers of these fields 

\begin{eqnarray}
&&\epsilon\,[A^\mu,B^{\mu\nu},b,d,\bar d,\psi^\ast,
\bar\psi^\ast,f^\ast,\bar f^\ast,e^\mu,c^\ast,\bar c^\ast,
d^\ast_\mu,\bar d^{*\mu},\eta]=0
\nonumber\\
&&\epsilon\,[\psi,\bar\psi,A^\ast_\mu,B^\ast_{\mu\nu},
b^\ast,d^\ast,\bar d^\ast,f,\bar f,e^\ast_\mu, 
c,\bar c,d^\mu,\bar d_{\mu},\eta^\ast]=1
\label{3.3}\\
\nonumber\\
&&{\rm gh}\,(d^\ast)=-3
\nonumber\\
&&{\rm gh}\,(c^\ast,d^\ast_\mu,\bar d,f^\ast)=-2
\nonumber\\
&&{\rm gh}(A^\ast_\mu,B^\ast_{\mu\nu},\psi^\ast,
\bar\psi^\ast,\bar c,b^\ast,\bar d_\mu,
e^\ast_\mu,\eta^\ast,\bar f)=-1
\nonumber\\
&&{\rm gh}(A^\mu,B^{\mu\nu},\psi,\bar\psi,
\bar c^\ast,b,\bar d^{*\mu},e^\mu,\eta,\bar f^\ast)=0
\nonumber\\
&&{\rm gh}(c,d^\mu,\bar d^\ast,f)=1
\nonumber\\
&&{\rm gh}(d)=2
\label{3.4}
\end{eqnarray}

\bigskip
The quantum theory is defined through the generating functional

\begin{equation}
\label{3.5}
Z_\Psi\,[J]=\int \prod[d\phi^A][d\phi^\ast_A]\,
\delta\,[\phi^{\ast}_A-{\delta\Psi\over\delta\phi^A}]\,
\exp\,{i\over\hbar}\left(W[\phi^A,\phi^\ast_A] 
+J_A\,\phi^A\right) 
\end{equation}

\bigskip\noindent 
where $\phi^A$ and $\phi^\ast_A$ respectively represent all the
fields and antifields appearing in (\ref{3.3}) and (\ref{3.4}) and
$W$ is a quantum action constructed starting from (\ref{3.1}). The
gauge fixing fermionic function can be chosen to be

\begin{eqnarray}
\Psi&=&-\int d^4x\,\Bigl[\bar c\,
\bigl(\partial_\mu\,A^\mu -{\alpha\over 2}\,b\bigl)
+\bar d_\mu\,\Bigl(\partial_\nu\,B^{\nu\mu}
-{\beta\over2}\,e^\mu\bigr)
\nonumber\\
&&\phantom{-\int d^4x\,\Bigl[\bar c\,
\bigl(\partial_\mu\,A^\mu -{\alpha\over 2}\,b\bigl)}
+\bar d\,\partial_\mu\,d^\mu
+\eta\,\partial^\mu\,\bar d_\mu\Bigr]
\label{3.2}
\end{eqnarray}

\bigskip \noindent
and the expectation value for an operator $X$ is given by 

\begin{equation}
<\,\, X \,\,>_{_{\Psi\,,J\,}}=\int\prod[d\phi^A]\,X\,
\exp\,\left({i\over\hbar}\,
W\,[\phi^A,\phi^\ast_A={\delta\Psi\over\delta\phi^A}]
+J_A\,\phi^A\right) 
\label{3.6}
\end{equation} 

\bigskip\noindent 
The condition that (\ref{3.5}) is independent of specific gauge
choices for null external sources, or equivalently, in the same
situation, that it must be invariant under  admissible changes in
$\Psi$, implies that the quantum master equation

\begin{equation}
<\,{1\over2}\,(W,W)-i\hbar\,\Delta W\,>_
{_{\Psi\,\,,J}}=0
\label{3.7}
\end{equation}

\bigskip\noindent 
must be satisfied. In Eq. (\ref{3.7}) the antibracket is defined as
$(X,Y)={\delta_rX\over\delta\phi^A}{\delta_lY\over\delta\phi^\ast_A}
-{\delta_rX\over \delta\phi^\ast_A}{\delta_lY\over\delta\phi^A}$ and 
the operator $\Delta$ as $\Delta\equiv{\delta_r\over\delta\phi^A}
{\delta_l\over\delta\phi^\ast_A}$.

\medskip
As can be observed, the operator $\Delta$ is potentially singular and
its action must be regularized. In this sense, the master equation at
loop order equal or greater than one is just formal unless a
regularization scheme is introduced. Expanding $W[\phi,\phi^\ast]$
in powers of $\hbar$ gives 
$W[\phi^A,\phi^{\ast}_A]=S[\phi^A,\phi^{\ast}_A]+
\sum_{p=1}^\infty\hbar^p M_p\,[\phi^A,\phi^{\ast}_A]$
and consequently the master equation (\ref{3.7}) can be written in
loop order. The first terms are

\begin{eqnarray}
\label{3.8}(S,S) &=& 0\\
\label{3.9}
(M_1,S) &=& \,i\, \Delta S
\end{eqnarray}

\bigskip
If we adopt a Pauli-Villars regularization with fermionic mass terms
and with the usual form for those of Dirac fields \cite{TNP}, it is
not difficult to show that the action of the $\Delta$ operator on $S$
is trivial, and so the theory is anomalous free.  This result is not
surprising because we know that $QED_4$ is anomalous free and also
because the actual theory has a fermionic covariant derivative that
reduces to that one of $QED_4$ under the correspondence (\ref{2.5}).
Since there is no obstruction of gauge symmetries, $W$ and $S$ can be
identified and we can integrate over the antifields to obtain the
gauge fixed version of (\ref{3.1}) as an effective action:

\begin{eqnarray}
\bar S&=&S_0+\int d^4x\,
\Bigl[i\,\partial_\mu\bar c\,\partial^\mu c
+\bigl(\partial_\mu A^\mu-{\alpha\over2}\,b\bigr)\,b
\nonumber\\
&&\phantom{S_0+\int d^4x\,
\Bigl[i\,\partial_\mu\bar c\,\partial^\mu c}
+i\,\partial\,_{[\mu}\bar d_{\nu]}\,\partial^{\mu} d^{\nu}
+\partial_\mu\bar d\,\partial^\mu d
\nonumber\\
&&\phantom{S_0+\int d^4x\,
\Bigl[i\,\partial_\mu\bar c\,\partial^\mu c}
+\bigl(\partial_\nu B^{\nu\mu}-{\beta\over2}\,e^\mu
-\partial^\mu\eta\bigr)\,e_\mu
\nonumber\\
&&\phantom{S_0+\int d^4x\,
\Bigl[i\,\partial_\mu\bar c\,\partial^\mu c}
-i\,\partial_\mu d^\mu\bar f + i\,f\partial^\mu\bar d_\mu\Bigr]
\label{3.10}
\end{eqnarray}

\bigskip\noindent 
If now we integrate out the tensor degrees of freedom, it is not
difficult to see that the effective action obtained in \cite{Barc1}
is generalized to

\begin{equation}
\bar S=\int d^4x\,\Bigl[{1\over2}\,\bar A_\mu\,
\bigl(\Box - m^2\bigr)\,\bar A^\mu
-\,{1\over2}\,\partial_\mu\, \bar A^\mu\,
\Bigl(1-{1\over\alpha} -{m^2\over\Box}\Bigr)\,
\partial_\nu\, \bar A^\nu\Bigr] + S_{ghost}
\label{effective}
\end{equation}

\bigskip\noindent 
where $\bar A_\mu=A_\mu+\frac{i}{mM}\bar\psi\gamma_\mu\psi$ is
essentially the vector gauge field shifted by the vectorial current.
So it appears an effective mass term for the vector fields, as can be
read from the inverse of the operator appearing in (\ref{effective}),
but also current-current self interactions in the fermionic
(effective) sector.

\medskip
The situation becomes completely different if we consider chiral
couplings with $\tilde A_\mu$. First it is necessary to replace the
covariant derivative $D_\mu$ appearing in (\ref{2.3}) and (\ref{2.4})
by

\begin{equation}
\tilde D_\mu=\partial_\mu+ieP_+\tilde A_\mu\,\,,
\label{3.11}
\end{equation}

\bigskip\noindent 
where $P_\pm={1\over2}(1\pm\gamma_5)$. The gauge invariances
appearing in the vector and tensor sectors do not change, but we need
to consider the changes in the Dirac sector.  To do this, it is
enough to replace in Eq. (\ref{3.1})
$-ie\psi^\ast c\psi+ie\bar\psi c\bar\psi^\ast$ by
$-ie\psi^\ast P_-c\psi+ie\bar\psi cP_+\bar\psi^\ast$. The remaining
gauge fixing terms are not affected. The quantum master equation,
however, is not satisfied anymore. By using the same kind of
Pauli-Villars regularization previously adopted, we can see that
\cite{TNP} 

\begin{equation}
\Delta S_{Reg}=-\,{e^3\over{16\pi^2}}\int d^4x\,c\, 
\epsilon^{\mu\nu\rho\lambda}\,\tilde F_{\mu\nu}
\tilde F_{\rho\lambda}\,\,,
\label{3.12}
\end{equation}

\bigskip\noindent 
which is a similar expression to the one already developed in the
previous section. So we conclude that the $U(1)$ gauge symmetry is
obstructed if chiral couplings are present, but the contribution of
the tensor field to the anomaly is yet the one induced by
correspondence (\ref{2.5}). It is opportune to mention that the
nonrenormalizable vertex given by Eq. (\ref{2.3}) does not spoil the
result above because it does not enter in the triangle diagram.

\vspace{1cm}
\section{Conclusion}

\bigskip
We have shown that introducing fermions in a vector-tensor gauge
theory, coupled in a topological way with the tensor field, modifies
the quantum expression for the divergence of the axial Noether
current in a nontrivial manner, when compared to the usual vector
case.  The anomalous divergence expression, however, is yet a total
derivative, which is a consequence of the form of the chosen coupling
between fermions and the tensor field.

\medskip
It was also shown that although the proposed theory presents this
kind of anomalous current divergence, it does not present $U(1)$ or
tensor gauge anomalies. This fact permit us to integrate out the
tensor degrees of freedom, what generates mass for the vector sector.
It also introduces effective current-current couplings, but in such a
way that the $U(1)$ gauge symmetry is explicitly kept.

\medskip
We have also observed that when chiral couplings are permitted, true
gauge anomalies are generated, within a form that reduces to the
usual $FF^\ast$ one, once we redefine the gauge vector field by a
particular shift depending on the tensor fields.

\vspace{1cm}
\noindent
{\bf Acknowledgment:} This work is supported in part by Conselho
Nacional de Desenvolvimento Cient\'{\i}fico e Tecnol\'ogico - CNPq,
Financiadora de Estudos e Projetos - FINEP and Funda\c{c}\~ao
Universit\'aria Jos\'e Bonif\'acio - FUJB (Brazilian Research
Agencies).


\vspace{1cm}
\begin{thebibliography}{30}
\bibitem{Kalb} M. Kalb and P. Ramond, Phys. Rev. D9 (1974) 2273.
\bibitem{Cremmer} E. Cremmer and J. Scherk, Nucl. Phys. B72 (1974)
117. 
\bibitem{Allen} T.J. Allen, M.J. Bowick and A. Lahiri, Mod. Phys.
Lett. A6 (1991) 559.
\bibitem{Barc1} R. Amorim and J. Barcelos-Neto, Mod. Phys. Lett.
A10 (1995) 917. 
\bibitem{Barc2} See also, J. Barcelos-Neto and S. Rabello, Z. Phys.
C74 (1997) 715.
\bibitem{Green} See, for example, M.B. Green, J.H. Schwarz and E.
Witten, {\it Superstring theory} (Cambridge University Press, 1987)
and references therein
\bibitem{Dirac} P.A.M.  Dirac, Can. J. Math. 2 (1950) 129; {\it
Lectures on quantum mechanics} (Yeshiva University, New York, 1964).
\bibitem{Kaul} We mention among others, R.K. Kaul, Phys. Rev. D18
(1978) 1127; C.R. Hagen, Phys. Rev. D19 (1979) 2367; A. Lahiri, Mod.
Phys.  Lett. A8 (1993) 2403; J. Barcelos-Neto and M.B.D. Silva, Int.
J. Mod.  Phys. A10 (1995) 3759; Mod. Phys. Lett. A11 (1996) 515; R.
Banerjee and J. Barcelos-Neto, {\it Reducible constraints and the
phase space extension in the canonical formalism}, hep-th 9703020, to
appear in Ann. Phys.
\bibitem{Henneaux} See also M.  Henneaux and C. Teitelboim, {\it
Quantization of gauge systems} (Princeton University Press, New
Jersey, 1992) and references therein.
\bibitem{Clark} T.E. Clark and S.T. Love, Nucl. Phys. B223 (1983)
135; W. Bardeen and N. Deo, Nucl. Phys. B264 (1986) 364. See also, J.
Barcelos-Neto and A. Das, Mod. Phys. Lett. A5 (1990) 2573.
\bibitem{Fujikawa} K. Fujikawa, Phys. Rev. Lett. 42 (1979) 1195;
Phys. Rev. D21 (1980) 2848; 22 (1980) 1499 (E).
\bibitem{BV1} I. A. Batalin and G. A. Vilkovisky, Phys. Lett. B102
(1981) 27; I. A. Batalin and G. A. Vilkovisky, Phys. Rev. D28 (1983)
2567.
\bibitem{Gomis} J. Gomis, J. Paris and S. Samuel, Phys. Rep. C259
(1955) 1. 
\bibitem{TNP}  W.Troost, P.van Nieuwenhuizen and
A. Van Proeyen, Nucl. Phys. B333 (1990) 727.
\end{thebibliography}
\end{document}